\documentclass[11pt]{article}
\usepackage[utf8]{inputenc}
\usepackage[paper=a4paper,centering]{geometry}

\usepackage{graphicx}
\usepackage{amsmath}
\usepackage{amssymb}
\usepackage{setspace}
\usepackage{hyperref}
\usepackage{authblk}
\usepackage{xcolor}
\usepackage[nottoc]{tocbibind}
\usepackage{physics}

\usepackage[sorting=none]{biblatex}
\renewbibmacro{in:}{}
\title{How many particles do make a fluid? \\ Qualifying collective behavior in expanding ultracold gases}
\author[1]{Stefan Floerchinger\footnote{floerchinger@thphys.uni-heidelberg.de}}
\author[1]{Giuliano Giacalone\footnote{ giacalone@thphys.uni-heidelberg.de}}
\author[1]{Lars H. Heyen\footnote{heyen@thphys.uni-heidelberg.de}}
\author[1]{\\Leena Tharwat\footnote{leena.tharwat@stud.uni-heidelberg.de}}
\affil[1]{Institut f\"ur Theoretische Physik, Universit\"at Heidelberg, Philosophenweg 16, 69120 Heidelberg, Germany}

\begin{document}

%\date{\vspace{-30pt}}
%\date{}

\maketitle
\thispagestyle{empty}

\begin{abstract}
Collective phenomena in quantum many-body systems are often described in terms of hydrodynamics, an appropriate framework when the involved particle numbers are effectively macroscopic. We propose to use experiments on expanding clouds of few and many interacting cold atoms to investigate the emergence of hydrodynamics as a function of particle number. We consider gases confined in two-dimensional elliptically-deformed traps, and we employ the manifestation of elliptic flow as an indicator of collective behavior. We quantify the response of the gas to the deformation of the trapping potential, and show how such information can be used to establish how many atoms are needed for the system to develop a degree of collectivity comparable to that expected in the hydrodynamic limit. This method permits one, in particular, to exploit observations made in expanding atomic gases to shed light on the apparent hydrodynamic behaviour of mesoscopic systems of quarks and gluons formed in the scattering of light ions in high-energy collider experiments.
\end{abstract}

%\tableofcontents

\section{Introduction}

Emergent hydrodynamic behavior is observed in many-body quantum systems across a wide spectrum of energy scales, ranging from ultracold gases \cite{RevModPhys.80.885,RevModPhys.80.1215} ($T\approx10^{-6}$ K) created in tabletop experiments, to the hot and dense strong-interaction matter, the quark-gluon plasma \cite{Shuryak:2014zxa,Busza:2018rrf} ($T\approx 10^{12}$ K \cite{Gardim:2019xjs}), produced in nuclear collisions performed in the world's largest accelerator machines. Hydrodynamics is an effective description of the long-time and long-distance behavior of a system, based on conservation laws \cite{Schafer:2009dj,Schaefer:2014awa}.  A substance behaves hydrodynamically as soon as the relaxation of its conserved charges occurs on a time scale that is large compared to the microscopic time associated (typically) with the collisions of its constituents. 
%The response of conserved quantities to external perturbations, encoded in the equations of hydrodynamics, is inherently non-local and occurs via diffusion, or \textit{collective} motion.

Observing the collective response to a perturbation does not alone imply a fluid description. The connection between collective and hydrodynamic behavior has been strongly emphasized in the context of high-energy nuclear experiments since the beginning of the heavy-ion collision program. Observation of strong elliptic flow, $v_2$, in off-central nucleus-nucleus collisions \cite{Heinz:2013th} indicates that the quark-gluon plasma responds collectively to an elliptical deformation of its geometry. The response of the system to this deformation, quantified via the ratio $v_2/\varepsilon_2$ \cite{Voloshin:1999gs,Bhalerao:2005mm}, where $\varepsilon_2$ is the elliptic anisotropy of the quark-gluon plasma prior to its expansion, has then been employed to confront elliptic flow measurements to the predictions of ideal and viscous hydrodynamics, as well as transport theory \cite{Molnar:2004yh,NA49:2003njx,Drescher:2007uh,Alver:2010dn}. Analyses of this kind have been revamped in recent times \cite{Kurkela:2018qeb,Kurkela:2019kip,Kurkela:2020wwb,Roch:2020zdl,Kurkela:2021ctp,Ambrus:2021fej} due to the striking observation of sizable $v_2$ in so-called \textit{small systems} \cite{Nagle:2018nvi,Schenke:2021mxx}, involving either the collision of two light ions (e.g. protons), or the collision of a proton with a nucleus. Such systems possess no evident separation of scales that would justify a hydrodynamic picture. Collective behavior is however observed to persist down to collisions that emit as few as 20 particles per unit rapidity. Finding an interpretation for such observations is a pressing issue in the field. Do small system create a fluid-like quark-gluon plasma? Can fluid behavior manifest with such a few (albeit strongly-interacting) particles?

Motivated in part by these questions, in this paper we introduce an experimental method to study the type of collective behavior characterizing the expansion of cold atomic clouds. The question we want to address is precisely: how many (strongly-interacting) particles are needed for a system to develop an amount of collectivity close to that expected in the hydrodynamic limit? The response coefficient $v_2/\varepsilon_2$, not yet fully explored with expanding ultracold clouds, provides a straightforward tool to address this question.  The main conceptual advance brought by our analysis is the realization that this quantity can in fact be \textit{measured} in cold atom experiments, for any particle number. The first experimental determination of $v_2/\varepsilon_2$ will, thus, demonstrate a new method to characterize the collective behavior of few- and many-body quantum systems, bringing new insight on small systems in the context of high-energy experiments.

This manuscript is organized as follows. In Sec.~\ref{sec:2}, we review the principles of non-relativistic fluid dynamics, we introduce elliptic flow, and we discuss the response coefficient, $v_2/\varepsilon_2$, that is the main focus of this study. In Sec.~\ref{sec:3}, we review the status of elliptic flow measurements with cold atoms, and we explain our new proposed method to measure $v_2$ (and thus $v_2/\varepsilon_2$) for any particle number in such experiments. In Sec.~\ref{sec:4} we compute, then, the baseline $v_2/\varepsilon_2$ expected in quantum mechanics for a non-interacting Fermi or Bose gas, and we provide qualitative expectations for the outcome of experiments with interacting gases. Section~\ref{sec:5} is left for conclusive remarks and an outlook on further opportunities offered by our proposal.

\section{Elliptic flow as an emergent collective phenomenon}

\label{sec:2}

\subsection{Principles of fluid dynamics}

Before introducing the notion of elliptic flow, it is worth recalling the generic principles underlying the equations of non-relativistic fluid dynamics, and the approximations on which they are based. A similar review can be found in e.g. \cite{book,Schaefer:2014awa}. Hydrodynamics is a prime example of collective description for systems that are in local thermal equilibrium and that can be described by emergent macroscopic fields (e.g. velocity, temperature, chemical potential), in the sense of statistical physics. The dynamics of a system we classify as a fluid follows entirely from conservation laws of energy, mass (or particle number) and momentum. The conserved charges, are, hence, the mass density $\rho$, the momentum density $\mathcal{P}_k$, and the energy density, $\mathcal{E}$, which satisfy continuity relations,
\begin{align}
\nonumber \partial_t \rho+ \partial_j \rho_j &= 0,  \\ %\label{eq:MassConservation} 
\nonumber  \partial_t \mathcal{E}+ \partial_j \mathcal{E}_j&=0,  \\ %\label{eq:EnergyConservation}
\label{eq:MomentumConservation} \partial_t \mathcal{P}_{k}+ \partial_j \mathcal{P}_{jk}&=0,
\end{align}
where $j$ and $k$ label spatial coordinates, $\rho_j$ represents the mass current, $\mathcal{E}_j$ the energy current, and $\mathcal{P}_{jk}$ is the momentum density current. The local fluid velocity, $v_j$, is defined by the equation for the mass current,
\begin{equation}
    \rho_j= \rho v_j,
\end{equation}
and correctly transforms as a velocity under Galilei transformations. Splitting all quantities into a velocity dependent part and a part that transforms as a scalar with respect to Galilei boosts,
\begin{align}
\nonumber \mathcal{P}_{k}=\rho v_k, \hspace{30pt} &\mathcal{P}_{jk}=\rho v_jv_k + T_{jk} \label{p},\\ 
\mathcal{E}= \frac{1}{2} \rho v^2+\epsilon , \hspace{30pt} &\mathcal{E}_j= \mathcal{E} v_j+ v_iT_{ij}+q_{j},
\end{align}
we have introduced an internal energy density $\epsilon$, the stress tensor, $T_{jk}$, which is symmetric in absence of an external torque, and the heat flow $q_j$. Consequently, Eqs.~\ref{eq:MomentumConservation} become
\begin{align}
\nonumber    &\partial_t \rho + \partial_j (\rho v_j) = 0 , \\ %\label{eq:MassConservation2} \\
\nonumber    &\rho(\partial_t + v_j \partial_j) v_k + \partial_j T_{jk} = 0 , \\ %\label{eq:MomentumConservation2} \\
    &(\partial_t + v_j \partial_j) \epsilon + \epsilon \partial_j v_j + (\partial_j v_k) T_{jk} + \partial_j q_j = 0. \label{eq:EnergyConservation2} 
\end{align}
We obtain, thus, a system of 14 variables but only 5 independent equations to determine them.

To move further, one exploits the fact that hydrodynamics is a description of long-wavelength properties that vary slowly in time, such that it is natural to consider a systematic expansion in the derivatives of the thermodynamic fields $\rho$, $v_j$, $\epsilon$\footnote{Other fields can be chosen, e.g., temperature, $T$, and chemical potential, $\mu$, which are related to $\rho$ and $\epsilon$ via the equation of state $p(\mu,T)$, namely, $\rho(\mu,T)=m\frac{\partial p}{\partial \mu}$, and $\epsilon(\mu,T)=-p(\mu,T)+T\frac{\partial p}{\partial T} + \mu \frac{\partial p}{\partial \mu}$.}. The derivative expansion does not converge, and one has to truncate it.  To zeroth order in the fluid velocity, the stress tensor reads
\begin{equation}
    T_{jk} = \delta_{jk}p~,
\end{equation}
corresponding to ideal fluid dynamics. The second line of Eq.~(\ref{eq:EnergyConservation2}) for the conservation of momentum becomes, in this case, the Euler equation
\begin{equation} 
\label{Euler}
\rho(\partial_t + v_j\partial_j)v_k = - \partial_k p~,
\end{equation}
where $(\partial_t + v_j\partial_j)$ is also referred to as the \textit{material} derivative. The first order correction to the stress tensor, involving one power of gradients, has the following form
\begin{equation}
\label{eq:Tjk_1st}
    T_{jk}= \delta_{jk}p-(2\eta\sigma_{jk}+\delta_{jk}\zeta\partial_i v_i)~, 
\end{equation}
where we have now introduced the shear stress tensor $\sigma_{jk}= \frac{1}{2}\partial_j v_k+ \frac{1}{2}\partial_k v_j - \frac{1}{3}\delta_{jk}\partial_i v_i$, which is the only traceless and symmetric tensor of first order in the derivatives of $\rho, \epsilon$ and $v_j$. The (positive and dimensionful) coefficients of the linear combination in Eq.~(\ref{eq:Tjk_1st}), $\eta$ and $\zeta$, are, respectively, the shear viscosity and the bulk viscosity of the fluid. The conservation of momentum leads, then, to the Navier-Stokes equation,
\begin{equation} 
\label{eq:NavierStokes}
\rho(\partial_t + v_j \partial_j) v_k = - \partial_k p + \biggl \{ \partial_j \bigg[\eta\big(\partial_j v_k + \partial_k v_j - \frac{2}{3} \delta_{jk} \partial_i v_i\big)\bigg] - \partial_k [\zeta \partial_i v_i] \biggr\}.
\end{equation}
Similarly, in the equation for the conservation of energy, the last line in Eq.~(\ref{eq:EnergyConservation2}), the first-order truncation of the heat flow involves a temperature gradient multiplied by an additional transport coefficient, $ q_j = -\kappa \partial_j T$, where $\kappa$ is the thermal conductivity. Note that more transport coefficients appear if one goes beyond first order in the derivative expansion, although we do not consider this possibility here. The reduction of variables due to truncating the gradient expansion leads, then, to a system of equations like (\ref{eq:EnergyConservation2}), which, supplemented with an equation of state, and with the knowledge of the transport coefficients $\eta$, $\zeta$, and $\kappa$ for the considered substance, can now be solved (at least numerically).

%What is, then, the limit of validity of this description? The picture of the gradient expansion holds only if the viscous terms are small corrections on top of the ideal one. The relevant expansion parameter depends on the type of flow, which can either be compressible or incompressible. A compressible flow implies that the typical fluid velocity, $u$, is larger than the typical thermal velocity, such that the Mach number ${\rm Ma} = u / c_s$, where $c_s$ is the speed of sound defined by $c_s^2=\partial p / \partial \rho|_{s}$, is close to unity. Here we deal with expanding gases, for which the flow is compressible. Assuming for simplicity that the only nonvanishing transport coefficient is $\eta$, inspection of the Navier-Stokes equation shows that the magnitude of the viscous term relative to the ideal one is given by the so-called inverse Reynolds number, which gives the relevant expansion parameter:
What is, then, the limit of validity of this description? The picture of the gradient expansion holds only if the viscous terms are small corrections on top of the ideal one. The relevant expansion parameter depends on the type of flow, which can either be compressible or incompressible. A compressible flow implies that the typical fluid velocity, $v$, is larger than the typical speed of sound defined by $c_s^2=\partial p / \partial \rho|_{s/n}$, such that the Mach number 
\begin{equation}
    {\rm Ma} = v / c_s 
\end{equation}
is close to unity. Here we deal with expanding gases, for which the flow is compressible and Mach number could be of order unity. 
Assuming for simplicity that the only nonvanishing transport coefficient is $\eta$, inspection of the Navier-Stokes equation shows that the magnitude of the viscous term relative to the ideal one is quantified by the inverse Reynolds number, %which gives the relevant expansion parameter:
\begin{equation}
    {\rm Re}^{-1} = \frac{\eta}{\rho v L},
\end{equation}
where $L$ is the typical macroscopic length scale.
For situations where Ma is of order unity one can see $\text{Re}^{-1}$ as a parameter controlling the derivative expansion, while more generally this role is played by $\text{Ma}^2 \text{Re}^{-1}$ \cite{Ollitrault:2007du,Schaefer:2014awa}.
%the thermal velocity of the system. Hydrodynamics applies in the limit ${\rm Re}^{-1} \ll 1$. 
In kinetic theory it is more common to employ as expansion parameter the Knudsen number
\begin{equation}
    \text{Kn} = \frac{\text{Ma}}{\text{Re}} \sqrt{\frac{\gamma \pi}{2}} = \frac{\lambda_{\rm mfp}}{L},
\label{eq:definitionKnudsen}
\end{equation}
where $\gamma= c_p/c_V$ is the ratio of specific heats and $\lambda_{\rm mfp}=\eta/(\rho v)$ is the particle mean free path. The Knudsen number is a good expansion parameter in the hydrodynamic limit ($\lambda_{\rm mfp} \ll L$). 

We see, then, that both Reynolds and Knudsen numbers involve either transport coefficients or the particle mean free path. These quantities are encoded in the microscopic dynamics that underlies the fluid description, and as such they can be difficult to evaluate (e.g. in the case of quantum fluids). Knowing whether a system can be described by hydrodynamics or not is, thus, a nontrivial problem. This is especially true for \textit{mesoscopic} systems, which are the focus of the present manuscript. Our goal is to present a new method to assess whether the collective dynamics exhibited by generic few- and many-body systems resembles hydrodynamics, and to which extent it does so.

\subsection{Converting spatial anisotropy into momentum anisotropy}

\begin{figure}[t]
    \centering
    \includegraphics[width=\linewidth]{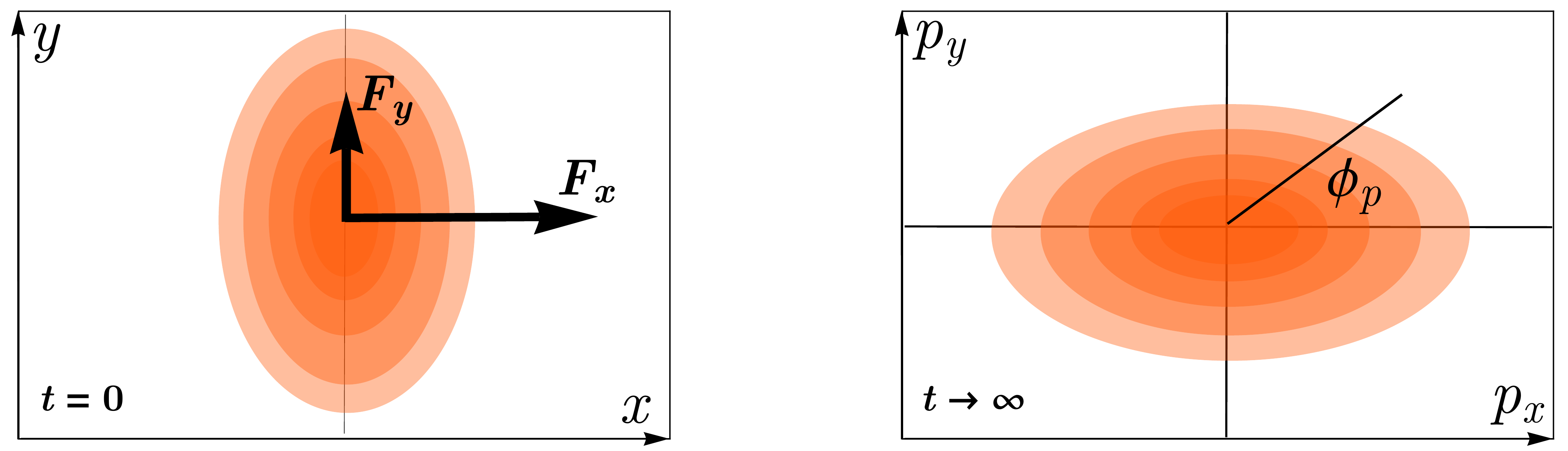}
    \caption{Illustration of the mechanism leading to elliptic flow in a fluid presenting an elliptical density profile. Left: initial condition with an elliptical asymmetry in the $(x,y)$ plane. Right: density of momentum at a later time during the hydrodynamic evolution. The distribution presents a $\cos(2\phi_p)$ modulation. Figure adapted from Ref.~\cite{Giacalone:2020ymy}.}
    \label{fig:1}
\end{figure}

We shall use the concept of elliptic flow, a straightforward consequence of the equations of hydrodynamics originally pointed out by Ollitrault \cite{PhysRevD.46.229}. We recall the Euler equation (\ref{Euler}), stating that the acceleration within the fluid is governed by a pressure gradient force (where $dv_j/dt$ is the material derivative of $v_j$)
\begin{equation}
\label{eq:euleq}
    \rho \frac{dv_j}{dt} = -\partial_j p.
\end{equation}
We follow the illustration of Fig.~\ref{fig:1} for a fluid in two dimensions. The left panel gives the state of a fluid with an elliptical deformation, elongated along $y$, at the initial condition. If we let such a system expand, according to Eq.~(\ref{eq:euleq}) the fluid experiences more acceleration along the $x$ direction than along $y$. Asymptotically, this imbalance of forces acting on the initial condition leads to a momentum distribution within the fluid that is elongated along $x$, as shown in the right panel of Fig.~\ref{fig:1}. By labeling $\phi_p$ the azimuthal momentum angle, the system presents in particular a $\cos (2\phi_p)$ asymmetry.

Elliptic flow is, hence, the quadrupole asymmetry of the asymptotic momentum distribution of flowing particles resulting from this mechanism of shape inversion. Following the notation of particle physicists, we dub $dN/d\phi_p$ the distribution of the azimuthal angles (in momentum space) of the detected particles (a quantity that integrates to $N$, the total number of particles). Elliptic flow is defined as the second Fourier harmonic of this spectrum:
\begin{equation}
    V_2 = \frac{1}{2\pi N} \int_0^{2\pi} \dd{\phi_p} \frac{dN}{d\phi_p} e^{i2\phi_p} \neq 0.
\end{equation}
Note that this is a complex number, but can be reduced to a single magnitude, $v_2\equiv |V_2|$, if the initial deformation of the fluid has a known orientation. It is further worth pointing out that if one relaxes the ideal fluid assumption of Eq~(\ref{Euler}), and considers the Navier-Stokes equation in Eq.~(\ref{eq:NavierStokes}),
%\begin{equation}
%\label{eq:NSfull}
%    \rho \frac{dv}{dt} = - \partial_j p  + \biggl \{ \partial_j \bigg[\eta\big(\partial_j v_k + \partial_k v_j - \frac{2}{3} \delta_{jk} \partial_i v_i\big)\bigg] - \partial_k [\zeta \partial_i v_i] \biggr\},
%\end{equation}
the second term on the right-hand side contributes to the force with a sign which is opposite to that of the pressure gradient, thus acting against the development of elliptic flow during the expansion of the fluid. This explains why elliptic flow provides a natural means to probe the viscous coefficients, $\eta$ an $\zeta$, of an expanding fluid \cite{Schafer:2009dj,Heinz:2013th}.
Let us also remark here that a non-interacting ideal classical gas at some temperature $T>0$ would not show any elliptic flow because the momentum distribution would be isotropic.

Elliptic flow can be considered as a smoking-gun of collective behavior within a many-body system. However, its observation does not alone imply that a fluid dynamic description is valid. 
For example a particle system driven by a Boltzmann equation with some collision kernel will develop elliptic flow, albeit less efficiently than hydrodynamics. Our goal is to demonstrate that elliptic flow can indeed be used as an indicator of the amount of collectivity observed in a given system, and to determine whether such collectivity is close to that expected in a genuine hydrodynamic scenario. 

\subsection{Quantifying the response to the initial deformation}

To give a \textit{measure} of how collective a system is, we use the following idea originally introduced by Poskanzer and Voloshin \cite{Voloshin:1999gs} in the context of high-energy nucleus-nucleus collision experiments. Elliptic flow emerges as a response of the system to the global deformation of its geometry. If we quantify the initial spatial deformation via an eccentricity parameter, $\varepsilon_2$, then the ratio $v_2/\varepsilon_2$ quantifies how efficiently the dynamics of the expansion has converted the initial elliptical deformation in position space into an elliptical modulation of the particle distribution in momentum space. If one knows, then, what the value of $v_2/\varepsilon_2$ is in hydrodynamics, then one can quantify how close the expansion dynamics is to hydrodynamic. 

 In the context of relativistic collision experiments, any statement concerning the \textit{fluidity} of the produced quark-gluon plasma based on the $v_2/\varepsilon_2$ ratio is plagued by a sizable theoretical uncertainty on the value of $\varepsilon_2$. In full generality, the eccentricity can be defined as the normalized quadrupole moment of the density profile \cite{Teaney:2010vd}
\begin{equation}
\label{eq:E2full}
\mathcal{E}_2 = \varepsilon_2 e^{i2\psi_2} = - \frac{\int_{\bf x} n({\bf x}) |{\bf x}|^2 e^{i2\phi} }{\int_{\bf x} n({\bf x}) |{\bf x}|^2  },
\end{equation}
where $\psi_2$ is the direction along which the elliptical anisotropy is oriented, the minus sign on the right-hand side is a convention, ${\bf x}$ is a coordinate in the $(x,y)$ plane where the system lies, while $n({\bf x})$ represents the density of particles (sometimes energy density is used in the relativistic context). In heavy-ion collisions, the shape of the energy density of the quark-gluon plasma can not be resolved by the particle detectors, but can only be inferred indirectly via model assumptions, which engenders a significant uncertainty.

\begin{figure}[t]
    \centering
    \includegraphics[width=\linewidth]{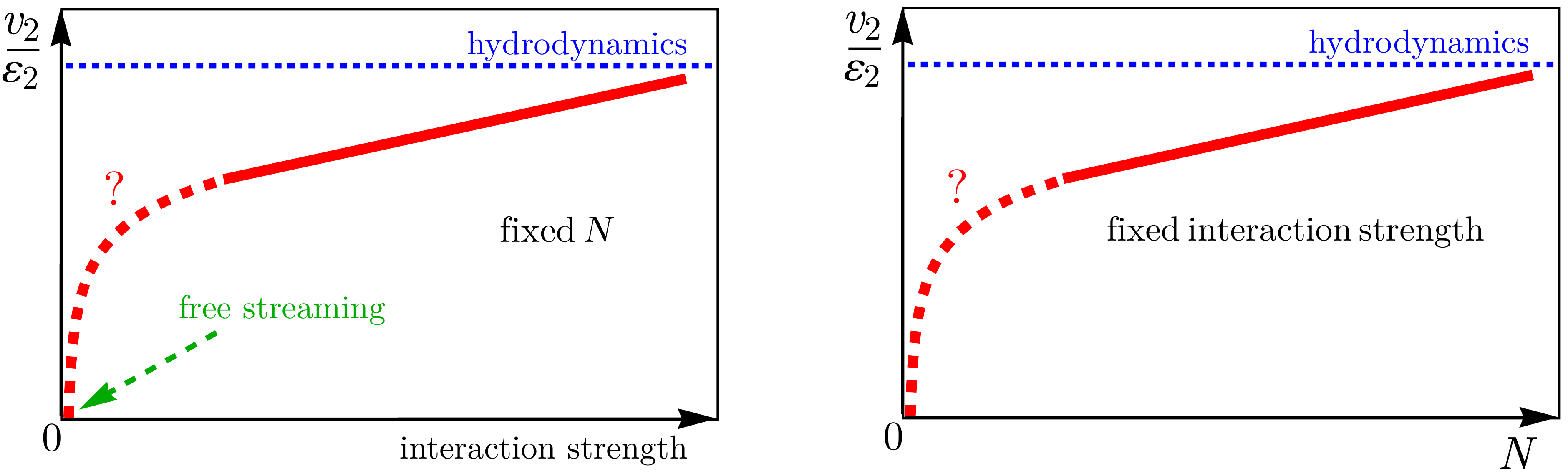}
    \caption{Left: Sketched qualitative evolution of $v_2/\varepsilon_2$ as a function of the inter-particle interaction strength, and fixed particle number, $N$. Right: same but at fixed interaction strength and varying particle number. The dashed parts of the proposed trajectories represent the regimes where quantum effects may play a decisive role. 
    %For simplicity, the proposed curves have the same shape in both panels, though this may not represent a realistic description of the physical system under consideration. The horizontal blue dashed lines represent hypothetical hydrodynamic limits.
    }
    \label{fig:2}
\end{figure}

This leads us to the most important point of this manuscript. In the context of experiments on expanding cold atomic clouds, the choice of $\varepsilon_2$ is part of the experimental setup, and the value of $v_2/\varepsilon_2$ can therefore be \textit{measured}. This opens up the possibility of performing quantitative studies of the nature of the collective behavior of expanding gases via measurements of elliptic flow. Here we are concerned with the onset of hydrodynamic behavior. This can be addressed in two main ways, illustrated in Fig.~\ref{fig:2}.\footnote{We discuss a picture where hydrodynamics emerges from interactions. One should note that elliptic flow is experimentally observed also in Fermi gases at zero temperature governed by superfluid hydrodynamic equations \cite{PhysRevLett.89.250402,doi:10.1126/science.1079107}, where the pressure gradient is not driven by collisional forces, but interaction terms still play a decisive role.} 

As done in the plot on the left of the figure, one can look at a sample with a given particle number, $N$, and tune the interactions among constituents. A non-interacting system undergoes a free streaming (or \textit{ballistic}) expansion, which does not develop any anisotropy.\footnote{This is not the case in a quantum gas, where, for non-interacting particles, quantum effects modify nontrivially $v_2/\varepsilon_2$ in the few-body limit. We will address this in Sec.~\ref{sec:4}.} 
Increasing the interaction strength, the ratio $v_2/\varepsilon_2$ converges to its hydrodynamic value.
Alternatively, as illustrated in the right-hand plot of Fig.~\ref{fig:2}, one can keep the inter-particle interaction strength fixed, and vary the number of particles. 
Hydrodynamics is then recovered for $N\rightarrow\infty$. 
We note that the trajectory of $v_2/\varepsilon_2$ for a \textit{classical} gas would smoothly and monotonically approach zero in both panels of Fig.`\ref{fig:2}, i.e., both in the limit both low particle numbers or weak interactions (the dashed regions of the proposed curves). 
For a \textit{quantum} gas, such regimes are more difficult to predict, and require experimental investigations.

Both scenarios presented in Fig.~\ref{fig:2} can be explored by means of cold atom experiments. First of all, in analogy with the situation realized in the context of high-energy nuclear collisions (and discussed in Fig.~\ref{fig:1}), one can prepare cold clouds lying in effectively elliptically-deformed two-dimensional shapes. Atoms are typically trapped in harmonic oscillator potentials characterized by frequencies $\omega_{x}$, $\omega_y$ and $\omega_z$, such that an effective 2D system can be obtained by setting one of these frequencies to a value much higher than the others. Secondly, the inter-atom interaction strength can be tuned experimentally. In a Fermi gas, the strength of interaction can be quantified via the product $k_F a$ \cite{book2}, where $a$ is the magnitude of the $s$-wave scattering length of the two-body interactions, while $k_F$ is the Fermi wave number, giving the typical inter-particle distance, $k_F\propto n^{1/D}$ (for $D$ spatial dimensions). At a given $N$, the value of $a$, and thus the interaction strength, can be magnetically tuned by means of Feshbach resonances \cite{RevModPhys.82.1225}. Finally, optical techniques to prepare high-fidelity states down to $N=1$ in two-dimensional systems are also available \cite{PhysRevA.97.063613}. The curves proposed in Fig.~\ref{fig:2} could, thus, be readily investigated experimentally with expanding ultracold gases.\footnote{Let us briefly comment that similar studies may also be conducted in the three-dimensional case.
While in general more complex, these systems can in fact provide additional symmetry. 
This is the case for a 3D gas at a strongly interacting fixed point of the atomic interaction (saturating the unitary bound), which leads in particular to conformal invariance.
%This fixed point in two spatial dimensions coincides with the non-interacting case (which always has the conformal symmetry). 
In two spatial dimensions there is no strongly interacting fixed point, but only the non-interacting one, which also has a conformal symmetry. 
In this case, symmetry allows for analytical calculations of certain isotropic quantities and access to anisotropic quantities in a perturbative expansion in the anisotropy, even in the presence of strong interactions \cite{PhysRevA.100.023601}.}

In this paper, we are mostly concerned with the case where interactions are kept fixed and the particle number is varied. As we discuss in the next section, this is, on the one hand, the kind of setup has not been explored yet in studies of elliptic flow with expanding ultracold gases, and on the other hand the case that most naturally connects to the situation of relativistic nuclear collisions, as one moves from large to small collision systems. We have, thus, outlined a well-defined conceptual method to exploit elliptic flow in the classification of the collective behavior of few- and many-body systems. In the remainder of this manuscript, we explain how $v_2/\varepsilon_2$ can be measured in practice for any value of $N$, and provide our predictions and expectations for the outcome of future experimental investigations.

\section{Measuring elliptic flow in expanding cold gases}

\label{sec:3}

\subsection{Studies at high particle number}

Before moving on to the description of elliptic flow in few- and many-body systems, let us briefly recall existing experimental investigations in the context of ultracold gases.  
The possibility that an ultracold atomic gas may develop elliptic flow has been first pointed out by Dalfovo \textit{et al.} \cite{RevModPhys.71.463}.
A theory of such phenomena, dealing with the expansion of a cloud of weakly-interacting Fermi atoms at zero temperature driven by superfluid hydrodynamic equations has been presented shortly afterwards in a remarkable paper by Menotti \textit{et al.} \cite{PhysRevLett.89.250402}. The first experimental observation of elliptic flow has been obtained at Duke university by O'Hara \textit{et al.} \cite{doi:10.1126/science.1079107}, whose results we shall now analyze in detail.  

In this experiment, a cloud of strongly-interacting ($k_F a \gg 1$) $^6$Li atoms is squeezed into a two-dimensional trap with a strong elliptical asymmetry. The aspect ratio of the cloud is defined by
\begin{equation}
    \lambda = \frac{\omega_y}{\omega_x},
\end{equation}
where $\omega_j$ is the frequency of the confining harmonic oscillator potential along direction $j$. We introduce, hence, the ellipticity of the trap, $\varepsilon_2$, via the following universal definition,
%\footnote{We note that Eq.~(\ref{eq:E2def}) is in fact consistent with the generic definition of the eccentricity given in Eq.~(\ref{eq:E2full}). For instance, if $n({\bf x})$ is a two-dimensional Gaussian of widths $\sigma_x$ and $\sigma_y$, then applications of Eq.~(\ref{eq:E2full}) yields $\varepsilon_2=\frac{1-\sigma_x^2/\sigma_y^2}{1+\sigma_x^2/\sigma_y^2}$, corresponding to Eq.~(\ref{eq:E2def}) if one identifies $\sigma_{x(y)}$ with $\omega_{x(y)}^{-1/2}$.}
\begin{equation}
\label{eq:E2def}
    \varepsilon_2 = \frac{1-\lambda}{1+\lambda}, 
\end{equation}
which satisfies $0<\varepsilon_2<1$ for $\omega_y < \omega_x$, and can be employed in any experimental setup without requiring the actual knowledge of the one-body particle density of the system [viz. $n({\bf x})$ in Eq.~(\ref{eq:E2full})].
However, for a single particle in the ground state of a two-dimensional harmonic oscillator, Eq. \eqref{eq:E2def} agrees in fact with Eq. \eqref{eq:E2full}.

\begin{figure}[t]
    \centering
    \includegraphics[width=.7\linewidth]{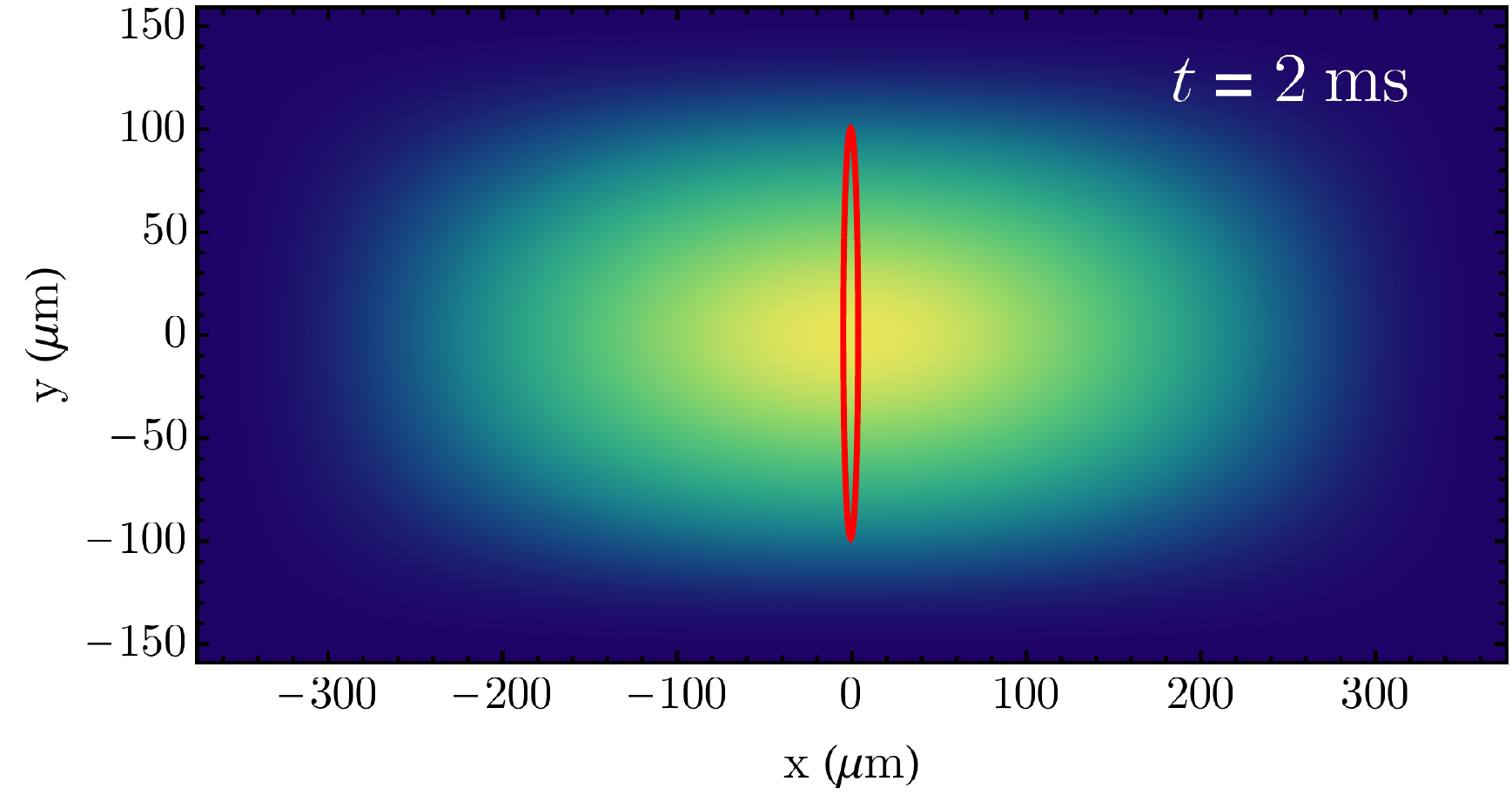}
    \caption{Shape of the cloud of ultracold $^6$Li atoms observed by O'Hara \textit{et al.} \cite{doi:10.1126/science.1079107} after an asymptotically large time $t=2$~ms. The profile is obtained from a two-dimensional Thomas-Fermi distribution of widths $\sigma_x=375$~$\mu$m and $\sigma_y=159$ $\mu$m. The red ellipse corresponds to the shape of the trap at the initial condition, where $\lambda=0.035$.}
    \label{fig:3}
\end{figure}

The experiment of Ref.~\cite{doi:10.1126/science.1079107} makes use of a trap that is as eccentric as experimentally possible:
\begin{equation}
    \lambda = 0.035~~ \rightarrow~~ \varepsilon_2 = 0.932 \, .
\end{equation}
Once the trap is released, the expansion of the cloud leads to an inversion of the initial shape and the observation of elliptic flow. The density of the cloud observed at the asymptotically large time $t=2$~ms is reproduced here in Fig.~\ref{fig:3}.\footnote{For sake of completeness, we note that the trapped cloud of O'Hara \textit{et al.} \cite{doi:10.1126/science.1079107} is not a genuine two-dimensional system, but rather a strongly-squeezed prolate ellipsoid with axial symmetry and a radial dimension much shorter than the axial one. The profile in Fig.~\ref{fig:3} represents, hence, a column-integrated particle density, though this is irrelevant for the present discussion.} The profile corresponds to a parameter-free two-dimensional Thomas-Fermi distribution~\cite{doi:10.1126/science.1079107}, with widths $\sigma_y=159$~$\mu$m, and $\sigma_x=375$~$\mu$m. The figure shows as well the shape of the trap at the initial condition, before release, as an ellipse. From the knowledge of the cloud profile, which we dub $\frac{dN}{dxdy}$, and assuming that at the asymptotic time the cloud has the same shape in both momentum and coordinate space, the value of elliptic flow can be then obtained directly from the profile in Fig.~\ref{fig:3} as
\begin{equation}
\label{eq:v2exp}
    v_2 = \langle \cos (2\phi) \rangle =  \int dx dy \frac{dN}{dxdy} \frac{x^2 - y^2}{x^2+y^2} = 0.42.
\end{equation}
Since $\varepsilon_2\approx1$, we further obtain
\begin{equation}
\label{eq:v2e2exp}
    v_2/\varepsilon_2 = 0.45,
\end{equation}
representing the first experimental determination of such quantity from an elliptic flow measurement. We note that this value is close to that found in model calculations of peripheral $^{208}$Pb+$^{208}$Pb collisions at the Large Hadron Collider \cite{Kurkela:2019kip}, where $v_2/\varepsilon_2\approx0.3$. A reason for this similarity is that the strongly-interacting Fermi gas at $T=0$ share similar magnitudes of viscous corrections with the quark-gluon plasma \cite{Teaney:2010vd}. In particular, while the values of $\eta$ for these two fluids differ by orders of magnitude due to the huge temperature difference, the quality of the fluid, $\eta/s$, where $s$ is the entropy density, is similar, $\eta/s\approx0.1$. In addition, we note that O'Hara \textit{et al.} \cite{doi:10.1126/science.1079107} repeat their experiments after switching off the interaction between atoms in the asymmetric trap. The disappearance of elliptic flow is observed for such a scenario, due to the fact that the critical velocity for the breakdown of superfluidity decreases with decreasing interaction strength, leading eventually to a ballistic expansion.

Subsequent experiments with expanding ultracold gases have investigated, among others, elliptic flow for both repulsive and attractive interactions \cite{PhysRevLett.90.230404,PhysRevLett.91.020402}, in the expansion of Fermi-Fermi mixtures \cite{PhysRevLett.106.115304}, as a probe of viscous corrections \cite{PhysRevLett.112.040405,PhysRevLett.113.020406,PhysRevLett.115.020401}, in the expansion of normal Bose systems \cite{PhysRevA.68.063603,PhysRevA.70.013607,PhysRevA.98.011601} and dipolar gases \cite{PhysRevLett.117.155301}. Of particular relevance for our analysis is the study by Fletcher \textit{et al.} \cite{PhysRevA.98.011601}. This experiment analyzes elliptic flow in an expanding cloud of $^{39}$K atoms for different choices of the interaction strength.  The idea is thus similar to that discussed in the left-hand side of Fig.~\ref{fig:2}, where $N$ is kept large.

Experimental studies of elliptic flow in cold atomic gases have a common denominator: they deal with regimes of very large (\textit{macroscopic}) numbers of atoms, where hydrodynamics becomes applicable as the system is strongly interacting. As anticipated, a systematic study of the disappearance of elliptic flow as one moves to \textit{mesoscopic} systems by reducing the number of trapped atoms, while keeping interactions active, is missing.  Data from proton-nucleons (or deuteron- and $^3$He-nucleus) collisions and proton-proton collisions at high energy points to the emergence of collective, hydrodynamic-like behavior in systems containing as few as 20 produced particles per unit rapidity \cite{CMS:2016fnw,ATLAS:2017rtr,PHENIX:2018lia,ALICE:2019zfl}.
It would be of paramount relevance to assess whether such behavior emerges as well in the expansion of cold atomic gases. We only need, then, a method to measure elliptic flow in expanding clouds of few atoms, where the notion of a continuous spectrum $\frac{dN}{dxdy}$ does not hold anymore. We establish now such a method. 

\subsection{Elliptic flow from few to many particles}

\begin{figure}[t]
    \centering
    \includegraphics[width=.8\linewidth]{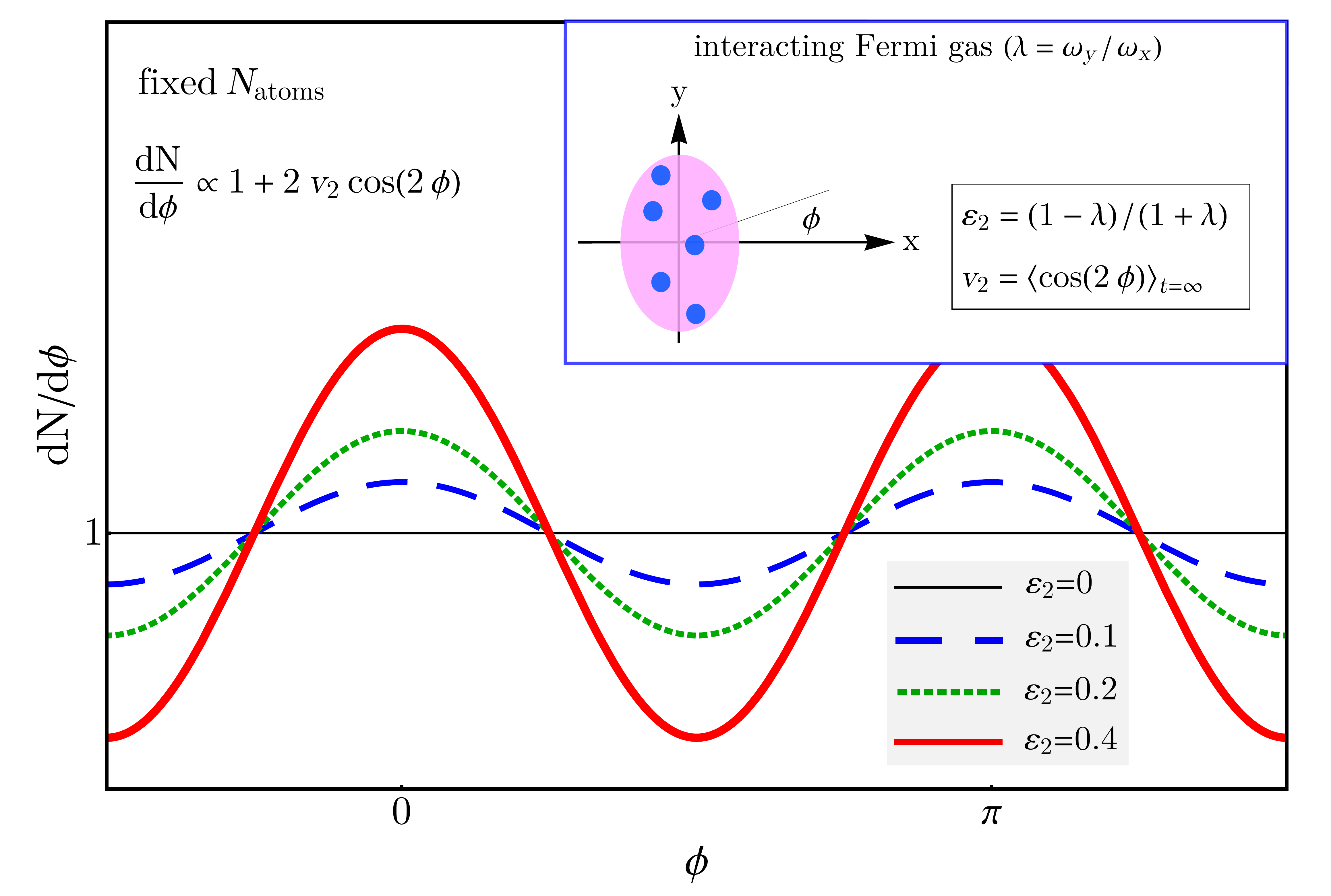}
    \caption{Illustration of the manifestation of elliptic flow in the asymptotic angular distribution of interacting atoms following the release of a trap presenting an elliptical asymmetry. The eccentricity of the trap, $\varepsilon_2$, is converted into a $\cos(2\phi)$ modulation of the particle spectrum, i.e., to nonzero elliptic flow, $v_2$, by the expansion. Different line styles correspond to different values for the initial $\varepsilon_2$. The inset shows an illustration of the initial condition of the system for $N_{\rm atoms}=6$.}
    \label{fig:4}
\end{figure}

In the hydrodynamic limit where $N\rightarrow\infty$, it is enough to perform the experiment once to measure $v_2/\varepsilon_2$. This is not the case in mesoscopic gases. At fixed $\varepsilon_2$, the expansion of a finite number of particles leads to a value of $v_2$ that fluctuates, with a statistical dispersion that depends on $N$. For such scenarios, one has to resort to a statistical description of $v_2$.  This leads to one of our main results, namely, the realization of a straightforward technique to achieve this goal. The orientation of the trap is known experimentally. If the short axis of the trap is along the $x$ direction, then
\begin{equation}
\label{eq:v2}
    v_2 = \langle \cos (2\phi_p) \rangle,
\end{equation}
where, as illustrated in Fig.~\ref{fig:4}, $\phi$ is defined such that $\phi=0$ (or $\phi=\pi$) is aligned with the $x$ direction, and the angular bracket denotes an average over the azimuthal angles of the atoms collected, asymptotically, in all the experiments. In other words, one should keep the orientation of the trap fixed, let the system expand, and then record the angles of the outgoing atoms. This has to be repeated a statistically significant number of times to reduce the dispersion of $\cos (2\phi)$. The outcome is a spectrum of azimuthal angles, $dN/d\phi$, which will look like that proposed in Fig.~\ref{fig:4}. If the systems responds collectively to its initial anisotropy by generating elliptic flow, then a $\cos( 2\phi)$ modulation emerges. This modulation becomes stronger when the trap ellipticity, $\varepsilon_2$, is increased.

A comment is in order, concerning the relation of this technique to that employed in measurements of $v_2$ in high-energy experiments. In heavy-ion collisions, the elliptical deformation of the quark-gluon plasma is mainly sourced by the nonzero distance (or \textit{impact parameter}) between the centers of the colliding nuclei. The \textit{orientation} of the resulting elliptical system is randomly fluctuating in the plane of the collision, and can not be controlled experimentally. Due to this, the average $\langle \cos (2\phi_p) \rangle$ vanishes. To overcome this problem, one has to compute instead $\langle \cos 2(\phi_1-\phi_2) \rangle$, where subscripts 1 and 2 label two distinct particles, and the average is now performed over all particle pairs in the selected sample of events. The distribution of the relative angle has the same structure as the histograms of Fig.~\ref{fig:4} in presence of elliptic flow. The drawback of this method is that the resulting $v_2$ contains all kind of spurious two-body correlations (dubbed in jargon \textit{non-flow}) emerging from phenomena unrelated to the dynamical response to the geometry of the system (coming e.g. from the decay of hadronic jets or high-mass particle resonances). Removing such effects to isolate the genuine collective signal requires highly nontrivial experimental effort. Cold atom experiments offer, thus, the unique possibility of controlling the direction of the ellipse\footnote{At high energy, measurements of particle emission with respect to a given axis are performed when the colliding objects are polarized. This is done in the measurement of transverse spin asymmetries \cite{Metz:2014bba} in either $p^\uparrow$-$p$ or $e^-$-$p^\uparrow$, where $p^\uparrow$ indicates a beam of polarized protons. A recent work \cite{Bozek:2018xzy} discusses the possibility of scattering polarized deuterons off large ions to exploit the experimentally-known orientation of the deuteron system and look for elliptic flow with respect to such a direction.}, and allow one to isolate elliptic flow from the simple one-body density in momentum space.

We have, thus, outlined a straightforward methodology to measure $v_2$, and consequently $v_2/\varepsilon_2$ in expanding mesoscopic cold atomic clouds. We can move on, then, to the presentation of our predictions and expectations for future experimental campaigns.

\section{Qualifying collectivity in expanding ultracold gases}

\label{sec:4}

In this section, we first present predictions for $v_2/\varepsilon_2$ as a function of atom number in the case of a non-interacting Fermi or Bose gas, which we are able to work out analytically. Subsequently, we discuss our expectations for interacting samples, and the emergent signatures of hydrodynamic collectivity driven by interactions. 

\subsection{Non-interacting Fermi gas}

Within a \textit{classical} picture, a system of non-interacting particles at some given temperature does not carry any momentum anisotropy, neither at the initial time, nor during or after the expansion. However, this is not the case in the limit of particles at low temperature, where quantum effects cannot be neglected.
As position and momentum representation of the wave function are linked by a Fourier transform, and the Fourier transform of an ellipse in position space is an ellipse rotated by 90$^\circ$ in momentum space,  if we set up one or more non-interacting particles in an anisotropic (elliptical) harmonic oscillator, we would expect the particles after release to move faster along the short axis. In this case, the collective behavior leading to the shape inversion is not driven by interactions, and can not qualify as hydrodynamic. This provides, essentially, a baseline for the magnitude of elliptic flow in absence of interactions at a given $N$. We consider now such an effect for gases at both $T=0$ and $T>0$, its magnitude compared to known hydrodynamic limits, and its scaling with the number of trapped fermions.
Subsequently, we extend our considerations to trapped bosons.

\subsubsection{Single particle, $N=1$, with minimal energy}

The time evolution of a single particle in a harmonic potential is governed by the Schrödinger equation (in units with $\hbar=1$),
\begin{equation}
    i \partial_t \ket{\psi} = \hat{H} \ket{\psi},
\end{equation}
where the Hamiltonian reads,
\begin{equation}
    \hat{H} = \frac{\hat{p}_x^2 + \hat{p}_y^2}{2m} + \frac{1}{2} m(\omega_x^2 \hat{x}^2 + \omega_y^2 \hat{y}^2) \, .
\end{equation}
In momentum space, this is solved by the well-known formula
\begin{align}
   \nonumber  \psi_{n_1, n_2}(p_x, p_y) &= \braket{p_x, p_y}{\psi_{n_1, n_2}} = \\ &\frac{1}{\sqrt{2^{n_1+n_2} n_1! \, n_2! \, \pi m \sqrt{\omega_x \omega_y}}} H_{n_1}\left(\frac{p_x}{\sqrt{m\omega_x}}\right) H_{n_2}\left(\frac{p_y}{\sqrt{m\omega_y}}\right) e^{-\frac{p_x^2}{2m\omega_x}-\frac{p_y^2}{2m\omega_y}} \, ,
\end{align}
where the $H_n$ are Hermite polynomials. 
%An isolated fermion can only occupy the ground state of the system. 
We assume now that the quantum particle has minimal energy and occupies the ground state.
The elliptic momentum anisotropy for this state can be calculated as 
\begin{equation}
\expval{\cos(2\phi_p)}_{\psi_{0,0}} = \expval{\frac{p_x^2 - p_y^2}{p_x^2 + p_y^2}}_{\psi_{0,0}} = \int \dd{p_x}\dd{p_y} \frac{p_x^2 - p_y^2}{p_x^2 + p_y^2} |\psi_{0,0}(p_x, p_y)|^2  = \frac{1-\sqrt{\lambda}}{1+\sqrt{\lambda}},
\end{equation}
which is a remarkably simple function of the aspect ratio, $\lambda = \omega_y/\omega_x$.

\begin{figure}[t]
    \centering
    \includegraphics[width=.5\linewidth]{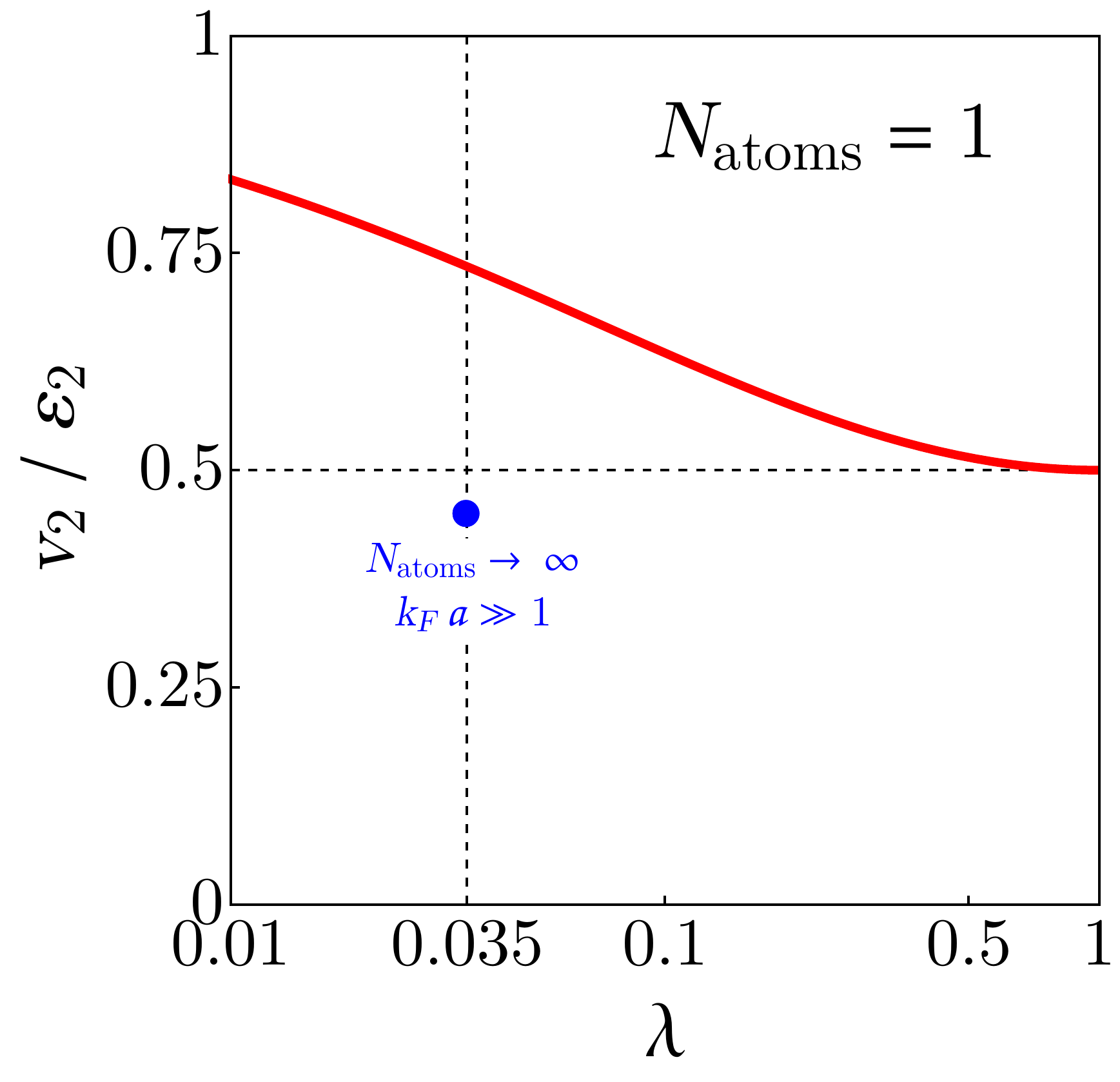}
    \caption{Prediction for $v_2/\varepsilon_2$ for the dynamics of a single fermion trapped in an anisotropic trap characterized by an aspect ratio $\lambda$. The eccentricity is defined as $\varepsilon_2=(1-\lambda^2)/(1+\lambda^2)$. The blue dot at $\lambda=0.035$ corresponds to the value of $v_2/\varepsilon_2$ measured by O'Hara \textit{et al.} \cite{doi:10.1126/science.1079107}}
    \label{fig:5}
\end{figure}

After turning off the trap, the particle expands freely. The free expansion does not influence the momentum anisotropy, as the operator $\cos(2\phi_p)$ commutes with the free Hamiltonian. Hence, the asymptotic momentum anisotropy is equal to the initial one
\begin{equation}
     v_2 = \frac{1-\sqrt{\lambda}}{1+\sqrt{\lambda}}.
     \label{eq:v2_oneparticle}
\end{equation}
The ratio $v_2/\varepsilon_2$, where we use $\varepsilon_2=(1-\lambda)/(1+\lambda)$ is plotted as a function of $\lambda$ in Fig.~\ref{fig:5}. It is equal to unity in the limit $\lambda=0$, and to 0.25 for $\lambda\rightarrow1$. With these conventions, we find, hence, that $v_2/\varepsilon_2$ is larger for a single particle than for the superfluid hydrodynamic system produced in the experiment of O'Hara \textit{et al.} \cite{doi:10.1126/science.1079107} in the hydrodynamic limit (with $N\sim10^5$). The curve in Fig.~\ref{fig:5} represents a new prediction of quantum mechanics to be tested with expanding cold atoms.

\subsubsection{Several particles, $N>1$, with minimal energy}

We move on, then, to the case of multiple non-interacting fermions in the same harmonic trap. At zero temperature these particles will fill up the energy levels with each level below the Fermi energy being occupied by exactly one particle (assuming a single spin polarization). For a non-interacting sample, the $N$-particle state factorizes into $N$ independent one-particle states (up to anti-symmetrization). Due to the fermionic nature of the particles, the state is anti-symmetrized as
\begin{equation}
    \ket{\Psi} = \frac{1}{\sqrt{N!}} \sum_{\sigma \in S_N} \text{sign}(\sigma) \bigotimes_{i = 1}^N \ket{\psi_{n_i, m_i}(\sigma(i))\,} \, ,
\end{equation}
where $\sigma$ is an element of the symmetric group $S_N$, and   $\ket{\psi_{n_i, m_i}(j)\,}$ is the one-particle state with excitation numbers $n_i$ and $m_i$ in $x$- and $y$-direction, referred to particle $j$.
In general, then, the expectation value of an operator $A_j=A(x_j, p_j)$, depending only on position and momentum of particle $j$, does not depend on the choice of such particle, i.e.,
\begin{equation}
    \expval{A_j}_\Psi = \frac{1}{N} \sum_{i=1}^N \expval{A}_{\psi_{n_i, m_i}},
    \label{eq:oneParticleOperatorEnsembleExpectationValue}
\end{equation}
due to the particles being indistinguishable. The expectation value with respect to $\ket{\Psi}$ of any operator can, hence, be written as an average:
\begin{equation}
    \expval{A}_\Psi = \frac{1}{N} \sum_{i=1}^N \expval{A_i}_{\Psi}.
    \label{eq:ensembleParticleOperatorEnsembleExpectationValue}
\end{equation}
With this notation in mind, it is instructive to evaluate the expectation value of $p_x^2$ (or, analogously, $p_y^2$), 
\begin{equation}
    \expval{p_x^2}_\Psi = m\omega_x \left( \frac{1}{2} + \frac{1}{N} \sum_{i=1}^N n_i \right).
\end{equation}
Assuming $\omega_y > \omega_x$, that means an excitation in the $y$ direction contributes more to the average momentum (squared) than one in the $x$ direction, but also that excitations in the $y$ direction require more energy. 
%A state with $m=0$ and one with $n=0$ that share the same total energy will also present $\expval{p_y^2}_{\psi_{0,m}} = \expval{p_x^2}_{\psi_{n,0}}$.
Since the fermions will fill up the energy levels in ascending energy order, we can estimate that similarly for a sufficiently high number of particles, the average momentum distribution will become isotropic. 

\begin{figure}[t]
    \centering
    \includegraphics[width=.9\linewidth]{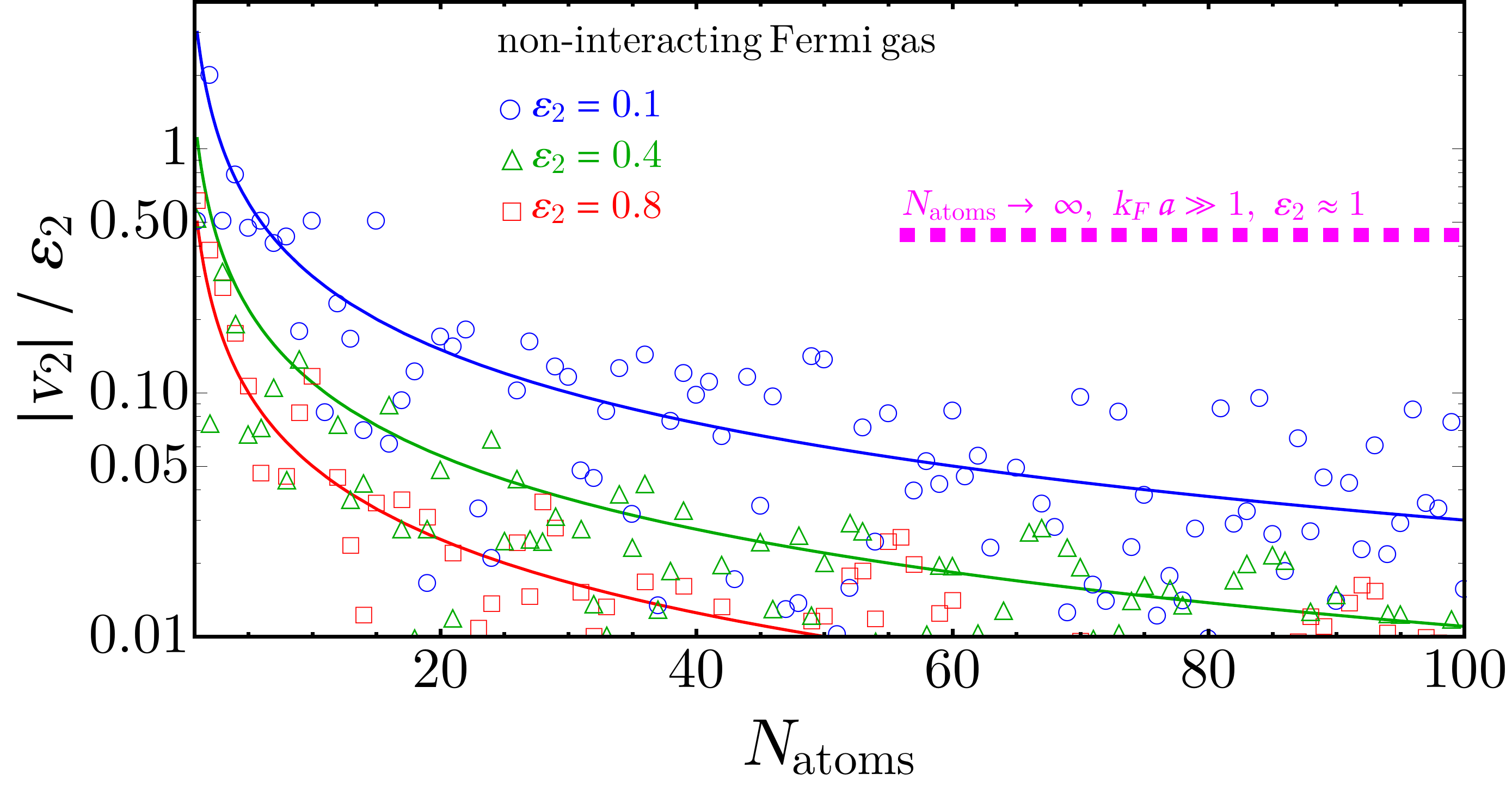}
    \caption{Predictions for $v_2/\varepsilon_2$ as a function of atom number for systems of $N$ non-interacting Fermions. Different symbols correspond to different choices of $\varepsilon_2$. The solid lines are proportional to $1/N$, and help guide the eye. The dashed line is the hydrodynamics result of O'Hara \textit{et al.} \cite{doi:10.1126/science.1079107}}
    \label{fig:6}
\end{figure}

As an example, a system of $N=2$ fermions would have two sets of quantum numbers, $(n_1, m_1)$ and $(n_2, m_2)$, which cannot be equal due to the anti-symmetrization.
For this simple system $v_2$ is then calculated as
\begin{equation}
    v_2 = \expval{\cos(2\phi_p)}_{\Psi_2} = \frac{1}{2} \left(\expval{\cos(2\phi_p)}_{\psi_{n_1,m_1}} + \expval{\cos(2\phi_p)}_{\psi_{n_2,m_2}} \right) \, .
\end{equation}
Calculating now $\langle \cos (2\phi_p) \rangle_\Psi$ numerically for different particle numbers, we obtain the ratio $v_2/\varepsilon_2$ that is plotted in Fig.~\ref{fig:6}. We indeed observe a dilution of $v_2$ with increasing $N$ which follows closely
\begin{equation}
    |v_2| \propto \frac{1}{N} \, .
\end{equation}
Increasing $\varepsilon_2$ leads to a further depletion of $v_2$, meaning that $v_2$ from quantum effects increases with $\epsilon_2$ weaker than linearly. We stress, once more, that the results of Fig.~\ref{fig:6} represent a prediction of quantum mechanics to be tested experimentally.
We emphasize that $v_2\rightarrow0$ for large $N$ is expected from simple considerations. As discussed e.g. in \cite{book2}, a non-interacting system of fermions described within the semi-classical picture of the Fermi-Dirac distribution undergoes ballistic-type expansion, and does not lead to any momentum anisotropy even if trapped in a deformed potential.  

\subsubsection{Finite temperature effects}

Going above zero temperature, the state of the system can no longer be defined as a single state $|\psi\rangle$, but must instead be described by a density operator $\rho$. In thermal equilibrium at temperature $T$ (or equivalently $\beta=1/(k_B T)$) and fixed particle number, the density operator reads
\begin{equation}
    \rho = \frac{e^{-\beta H^{(N)}}}{Z} = \frac{1}{Z}
    %\sum_{n_N > \dots > n_1}^\infty \sum_{m_N > \dots > m_1}^\infty 
    \sum_{(n_N, m_N) > \dots > (n_1, m_1) \in \mathbb{N}^2}
    |\Psi_{(n_1,m_1),\dots,(n_N,m_N)}\rangle \langle \Psi_{(n_1,m_1),\dots,(n_N,m_N)} | \, e^{-\beta \sum_{i=1}^N E_{n_i,m_i}} \, ,
\label{eq:densityOperator}
\end{equation}
where $H^{(N)}$ is the $N$-particle Hamiltonian, which in the non-interacting case discussed here is just the sum of the individual free Hamiltonians for each particle, $Z=\text{tr}(e^{-\beta H^{(N)}})$ is the partition sum, and lastly ``$>$'' is any ordering on $\mathbb{N}^2$ that avoids overcounting. Expectation values of operators with respect to this system are now given by
\begin{equation}
    \expval{A}_\rho = \text{tr}(\rho A) \, .
\end{equation}
With these definitions in mind we repeat the calculations from the previous subsections.

For a single particle, $N=1$, the calculation simplifies significantly since we can ignore the anti-symmetrization and the states only have two quantum numbers,
\begin{align}
    \nonumber \rho &= \frac{e^{-\beta H}}{Z} = \frac{1}{Z}\sum_{n=0}^\infty \sum_{m=0}^\infty |\psi_{n,m}\rangle \langle \psi_{n,m}| e^{-\beta (n \omega_x + m \omega_y)} \, , \\
    Z &= 
    %\sum_{n=0}^\infty \sum_{m=0}^\infty e^{-\beta (n \omega_x + m \omega_y)} = 
    \sum_{n=0}^\infty  e^{-\beta n \omega_x} \sum_{m=0}^\infty e^{-\beta m \omega_y} \, .
    \label{eq:oneParticleDensityOp}
\end{align}

\begin{figure}[t]
    \centering
    \includegraphics[width=.75\linewidth]{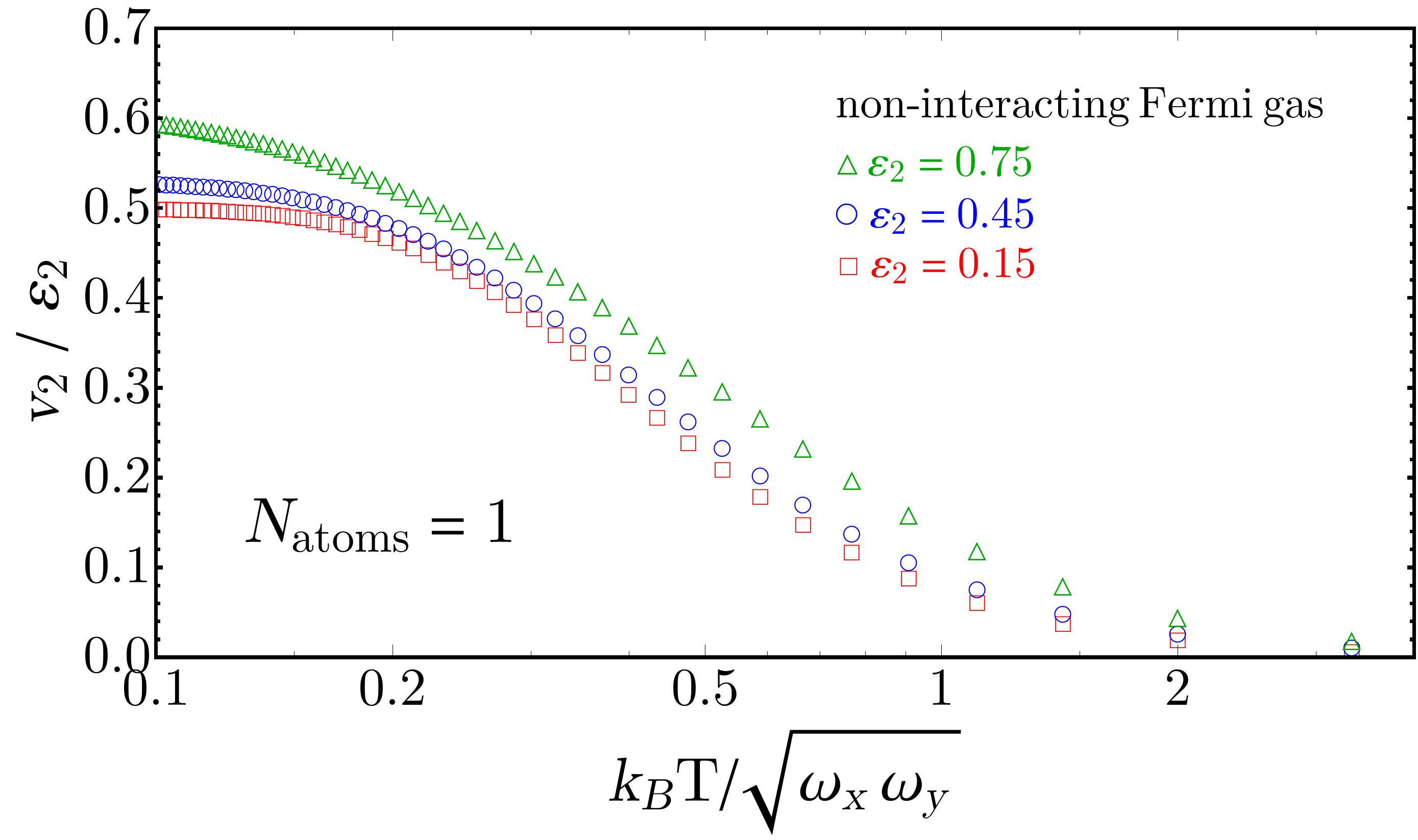}
    \caption{Temperature dependence of $v_2/\varepsilon_2$ for $N_{\rm atoms}=1$, obtained from Eq.~(\ref{eq:oneParticleDensityOp}) by summing $m$ and $n$ up to 400. Different symbols represent different values of $\varepsilon_2$. We note that the red curve in Fig.~\ref{fig:5} corresponds to the variation of $v_2/\varepsilon_2$ in the $T=0$ limit. We note that the combination $k_BT/\sqrt{\omega_x\omega_y}$ is dimensionless in units where $\hbar=1$.}
    \label{fig:7}
\end{figure}

With this we can now calculate expectation values for $\expval{p_x^2}$ (and analogously $\expval{p_y^2}$) analytically to be
\begin{equation}
\begin{split}
    \expval{p_x^2} &= 
    %\frac{1}{Z} \sum_{n=0}^\infty \sum_{m=0}^\infty \expval{p_x^2}_{n,m} e^{-\beta (n \omega_x + m \omega_y)} = m\omega_x \frac{\sum_{n=0}^\infty (n + \frac{1}{2}) e^{-\beta n \omega_x}}{\sum_{n=0}^\infty  e^{-\beta n \omega_x}} \\
    %&= m\omega_x \left( \frac{1}{2} + \partial_{\beta\omega_x} \text{ln}(1 - e^{-\beta \omega_x}) \right) = 
    m\omega_x \left( \frac{1}{2} + \frac{1}{e^{\beta \omega_x}-1} \right) \, .
\end{split}
\end{equation}
The ratio of the expectation values thus becomes
\begin{equation}
    \frac{\expval{p_y^2}}{\expval{p_x^2}} 
    %= \lambda \frac{1 + \frac{2}{e^{\beta\omega_y}-1}}{1 + \frac{2}{e^{\beta\omega_x}-1}} 
    = \lambda \frac{\text{tanh}(\beta \omega_x / 2)}{\text{tanh}(\beta \omega_y / 2)} \, .
\end{equation}
%with the dimensionless temperature $\tilde{T}= k_B T/\sqrt{\omega_x \omega_y}$.
In the low temperature limit, $\beta\rightarrow\infty$, the ratio of momenta converges to the aspect ratio of the trap frequencies, $\lambda,$ while in the high-temperature limit, $\beta\rightarrow 0$, the momentum aspect ratio approaches unity.
This suggests that $v_2$ should vanish as well at high temperatures, a behavior that is confirmed by our numerical calculations shown in Fig.\ref{fig:7}.
For this result we truncate the sums in Eq. \eqref{eq:densityOperator} at a cutoff $\Lambda_\text{num}=400$, which guarantees a good convergence of the calculated $v_2$ even for high temperatures, where higher energy excitations become more and more relevant.\footnote{Let us note that, as found in Fig.\ref{fig:5}, for $N_{\rm atoms}=1$ at $T=0$ the ratio  $v_2/\varepsilon_2$ grows monotonically with $\varepsilon_2$ (or with $1/\lambda$). This dependence is not modified by an increase of $T$ in Fig.~\ref{fig:7}. However, we have checked that this is not always the case. For $N>1$, the ordering of $v_2/\varepsilon_2$ with $\varepsilon_2$ can indeed be modified by a change in the temperature, which could also be investigated experimentally.}

For multiple particles, $N>1$, the cost of the numerical calculation grows exponentially in the number of particles (with simple algorithms), so a numerical study of these systems is beyond the scope of this paper. 
In general, pure quantum correlation effects driven mainly by the exclusion principle should disappear with increasing temperature. This is the case, for instance, for Pauli crystals \cite{PhysRevLett.126.020401}, which dissolve quickly as soon as the temperature of the system increases and thermal excitations become relevant ($k_B T \gg \sqrt{\omega_x\omega_y}$). 
As $v_2$ for a single particle vanishes at high $T$, it is natural to expect the same behavior to appear as well for higher numbers of particles, where $v_2$ is essentially given by an average of states. Numerically, we have checked that this is true up to $N_{\rm atoms}=4$. This generic expectation can also be justified by means of analytical considerations in the limit of very high temperature, i.e., that one has
\begin{equation}
    \lim_{T\rightarrow\infty} v_2 = 0 \, ,
\end{equation}
for any number of particles.
The relevant calculation can be found in appendix \ref{app:v2AtHighTemp}.

\subsubsection{Non-interacting bosonic gas}

Bosonic ensembles are symmetrized instead of anti-symmetrized. 
At $T=0$, a multiparticle state of non-interacting bosons thus has the form,
\begin{equation}
    \ket{\Psi} = \frac{1}{\sqrt{N!}} \sum_{\sigma \in S_N} \bigotimes_{i = 1}^N \ket{\psi_{n_i, m_i}(\sigma(i))\,}.
\end{equation}
To get to ensemble expectation values one can employ the same formulas as in the fermionic case, namely Eq.~\eqref{eq:oneParticleOperatorEnsembleExpectationValue} and Eq.~\eqref{eq:ensembleParticleOperatorEnsembleExpectationValue}.
However, as bosons occupy the same energy level, they are all in the ground state at minimal energy, such that the momentum anisotropy is given by the ground-state result, Eq.~\eqref{eq:v2_oneparticle}. 

For $T>0$, the calculation again requires a density operator.
The main difference with the fermionic case is that, due to the symmetrization, different particles having the same quantum numbers is allowed.
The computation of $v_2$ for any $N$ in such a scenario would thus proceed along the same lines of the previous subsection up to a redefinition of the populations of the considered states, and should lead to similar results in terms of scaling with $N_{\rm atoms}$. 
In addition, once the temperature is high enough that quantum effects become negligible, the bosonic and fermionic systems are expected to behave similarly.
In the limit $T\rightarrow\infty$, in particular, once can apply the result of Appendix \ref{app:v2AtHighTemp} to the bosonic case as well, such that we expect $v_2$ to vanish for any $N$.

%Finally, we note that for bosonic ensembles (which are symmetrized instead of antisymmetrized) once can employ the same formulas, namely Eq.~\eqref{eq:oneParticleOperatorEnsembleExpectationValue} and Eq.~\eqref{eq:ensembleParticleOperatorEnsembleExpectationValue}. However, as bosons occupy the same energy level, they are all in the ground state at zero temperature, such that the momentum anisotropy is given by the ground-state result, Eq.~\eqref{eq:v2_oneparticle}. 

\subsection{Expectations for interacting gases}

Quantitative estimates of $v_2$ as a function of the particle number in the mesoscopic regime require a detailed knowledge of the inter-atom interaction and the development of dedicated numerical tools, which is outside the scope of the present study. For higher values of $N_{\rm atoms}$, a kinetic description should become appropriate. Within such framework, the indicator of where in parameter space we might find the onset of fluid behavior is the Knudsen number defined in Eq. \eqref{eq:definitionKnudsen}.
%The smaller the Knudsen number (at least at large $N$), the more likely it is to find continuous flow instead of individual trajectories of (clusters of) particles.
In classical kinetic theory, we can find an expression for the mean free path in Eq. \eqref{eq:definitionKnudsen},
\begin{equation}
    \text{Kn} = \frac{1}{\sigma n L} \, ,
\end{equation}
where $L$ is the relevant macroscopic scale (the size of the atomic cloud), $\sigma$ is the particle-particle cross-section, and $n$ is the particle density.
We estimate now how these quantities scale with the number of particle number $N$.
The cross-section is a property of the interaction that does not depend on the particle number,
\begin{equation}
    \sigma \propto N^0 \, .
\end{equation}
The system size in $D$ spatial dimensions scales as
\begin{equation}
    L \propto \left( \sum_{i=1}^D \expval{x_i^2} \right)^{1/2} \propto N^{1/2} \, ,
\end{equation}
where the scaling $\expval{x_i^2} \propto N$ is an upper limit estimated from the one-dimensional harmonic oscillator.
The density will similarly scale like particle number per volume which becomes
\begin{equation}
    n \propto N L^{-D} \propto N^{D/2} \, .
\end{equation}
From these considerations, we obtain that the Knudsen number scales (at least) like
\begin{equation}
    \text{Kn} \propto N^{-(D+1)/2} \, ,
\end{equation}
which in two dimensions yields ${\rm Kn}\propto N^{-3/2}$. It decreases rather quickly with increasing particle number.
As for the dependence of Kn on interaction strength, that is very dependent on the spatial dimension.
In general the cross section grows with increasing strength of interactions, in our two-dimensional case roughly logarithmically \cite{2015arca.book....1L}.
So altogether the Knudsen number will decrease strongly with the number of particles and weakly with the interaction.
The (possibly) hydrodynamic regime, $\text{Kn} \ll 1$, we hence expect to reach rather quickly by increasing $N$ by the above heuristics.
In addition, the expansion of the cloud leads to an increase in system size and to a decline of the particle density, such that Kn will grow during the expansion of the atomic cloud.
This, at some point, will cause a hydrodynamic description to fail and be replaced by individual free particle trajectories (in analogy with the kinetic freeze-out of the quark-gluon plasma).
\begin{figure}[t]
    \centering
    \includegraphics[width=.75\linewidth]{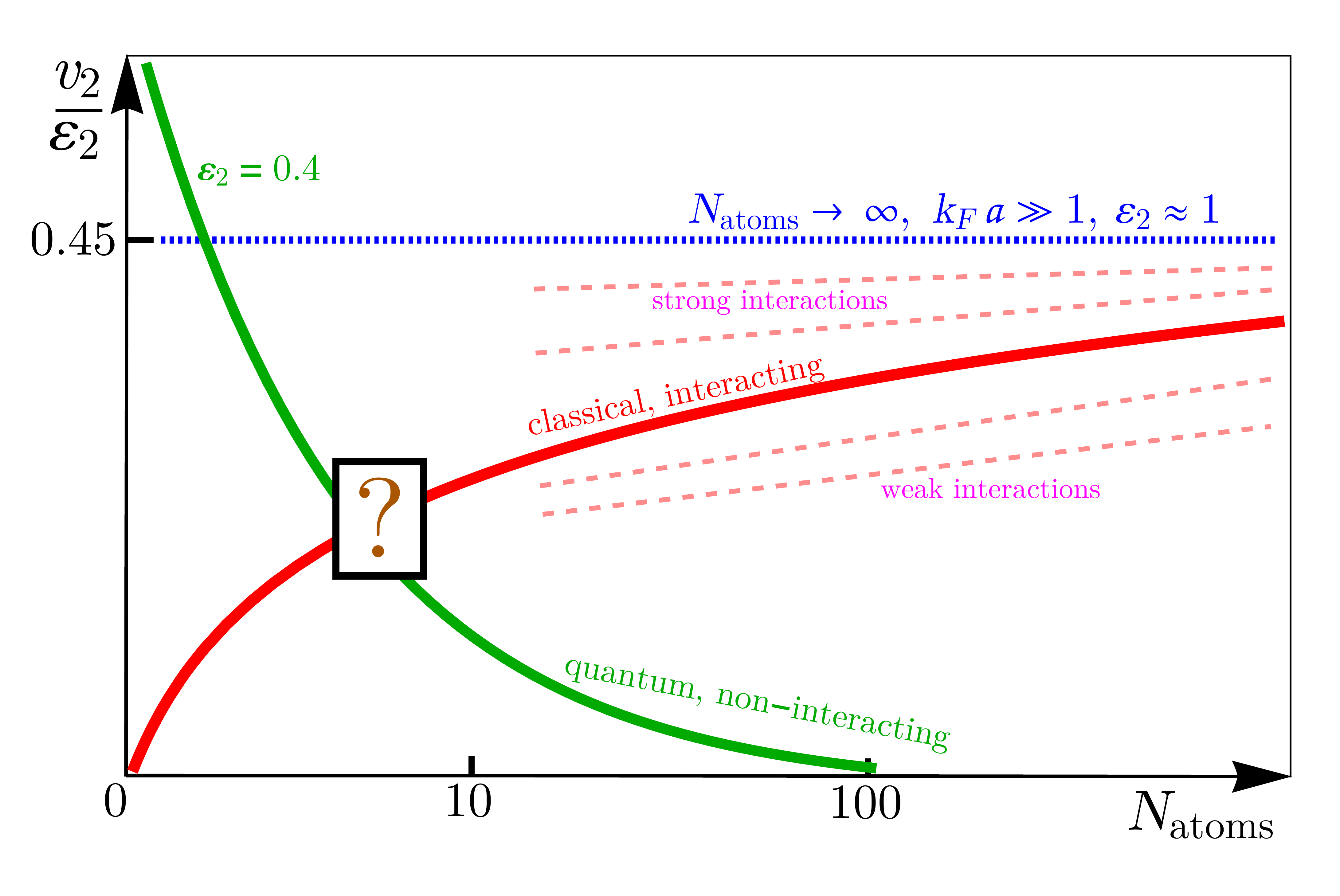}
    \caption{Expectation for the ratio $v_2/\varepsilon_2$ in the expansion of a cold atomic gas, as a function of the number of atoms. Pure quantum $v_2/\varepsilon_2$ is depicted as a solid line decreasing like $1/N$. Classical trajectories for an interacting gas start, on the other hand, at $v_2/\varepsilon_2=0$, and reach a maximum in the hydrodynamic limit, where $N_{\rm atoms}\approx10^3$ (for traps on the order of $\mu$m in size). The dotted line indicates the hydrodynamic limit obtained from the measurements of O'Hara \textit{et al.} \cite{doi:10.1126/science.1079107}. The precise way the system approaches hydrodynamics depends  on the interaction strength, as illustrated by the red dashed lines.}
    \label{fig:8}
\end{figure}

%Another very important condition for a fluiddynamic description working is the existence of local thermal equilibrium.
%For the system of trapped cold atoms, we can safely assume that the particles have had enough time to be in global thermal equilibrium before they are released.
%Throughout the following expansion this equilibrium will be maintained at least locally until the particles are too distant from each other.
%Keeping up the analogy to quark-gluon plasmas in heavy ion collisions, the moment in which the assumption of local equilibrium breaks down and the Knudsen number grows too large can be associated with the freeze-out surface.
%After that point the hydrodynamic description fails, but we can assume the system to be reasonable captured by individual free particle trajectories.

The global picture concerning the ratio $v_2/\varepsilon_2$ is given in Fig.~\ref{fig:7}, which is essentially a version of Fig.~\ref{fig:2} updated with all the insights gained from the previous discussions. The contribution of quantum effects, discussed in the previous subsection, is depicted as a green solid line that falls off like $1/N$ (very close to the case with $\varepsilon_2=0.4$ in Fig.~\ref{fig:6}). For an interacting classical system, as discussed in Fig.~\ref{fig:2}, the ratio $v_2/\varepsilon_2$ starts at zero and approaches the hydrodynamic limit for large $N_{\rm atoms}$.\footnote{Such number depends on the Fermi wave number as well as on the size of the trap, and for the typical setups of cold atom experiments, we expect that $10^3$ atoms is large enough.} For zero-temperature $^6$Li atoms and $\varepsilon_2\approx1$, such limit should correspond to that found by O'Hara \textit{et al.} \cite{doi:10.1126/science.1079107}, i.e., 0.45. In the mesoscopic regime, $N_{\rm atoms}\sim5$, both quantum and thermal excitations may contribute equally. For such region, we leave a question mark, to be investigated via experimental data.  Finally, once quantum effects should become subleading (maybe around $N\sim20$), the system approaches the hydrodynamic regime more or less steeply depending on the interaction strength, as illustrated by the dashed curves in Fig.~\ref{fig:7}.

Wrapping up, future experiments can map the onset of effects driven by interactions with increasing particle number, and investigate how quickly the ratio $v_2/\varepsilon_2$ converge to its hydrodynamic value. Hitting this hydrodynamic limit for few atoms would imply that an effective hydrodynamic description for such systems would lead to the right result, at least for this observable. Experimental measurements of the ratio $v_2/\varepsilon_2$ as a function of particle number would, thus, establish a new way of characterizing the collective dynamics of micro-, meso-, and macroscopic systems. This will open a new window onto the collective behavior of ultracold quantum gases, as well as shed new light on the apparent collectivity displayed by small systems in the context of high-energy collisions.

\section{Conclusion and outlook}

\label{sec:5}

We have discussed a method to characterize the collective dynamics of few- and many-body systems. It relies on the response, $v_2/\varepsilon_2$, of a considered system to an elliptical deformation of its geometry. We have pointed out, in particular, the unique possibilities offered by cold atom experiments in such kind of searches, namely, that $i)$ $v_2/\varepsilon_2$ can be measured experimentally, $ii)$ the number of particles in the system can be chosen precisely, $iii)$ the strength of inter-particle interactions be can be tuned. By studying how $v_2/\varepsilon_2$ approaches its hydrodynamic value, one can quantify how close the collective expansion exhibited by a given system is to that of a fluid.

There are several generalizations of our analysis that may lead to further experimental investigations. One could, for instance, deform the trap with a geometry more complicated than an ellipse. A triangular deformation would lead to so-called \textit{triangular flow} \cite{Alver:2010gr}, $v_3$, which can be measured experimentally in the same way as $v_2$ provided that the orientation of the triangle is known. One can, thus, look at the ratio $v_3/\varepsilon_3$ \cite{Alver:2010dn} to analyze the degree of collectivity. The same applies to all kinds of shapes \cite{Alver:2010dn,Kurkela:2020wwb}. In addition, one could search for the development of anisotropic flows in the case of isotropic traps. In such scenarios, the deformation of the system is determined from the random distribution of atom positions, which fluctuates randomly on a realization-by-realization basis due to density fluctuations. As we pointed out at the end of Sec.~\ref{sec:3}, techniques to measure anisotropic flow in absence of a globally-deformed geometry with a known orientation are commonly used in heavy-ion collisions, and could be readily adapted to cold atom experiments. Such measurements would, hence, provide a new tool to observe higher-order density-density correlations in trapped gases.

Studying the onset of fluid dynamic behavior with increasing particle number could also be conceptually interesting from a quantum information theoretic point of view.
Theoretical concepts like the generation of entanglement entropy and its relation to local dissipation \cite{DowlingFloerchingerHaas:2020} could be more tractable when particle numbers are small.

A potentially far-reaching extension of our study could finally involve the analysis of the collective behavior of systems that are not in local thermal equilibrium. This is unlike standard cold atom experiments where the systems are locally thermalized before the release of the trap. Mesoscopic out-of-equilibrium systems should not be hydrodynamic, however, they may still develop a sizable elliptic flow. If one were able to tune the degree of equilibration of a given gas, one could study the emergence of collective behavior via the ratio $v_2/\varepsilon_2$ as a function of the degree of thermalization. This is very close to the problem that one encounters when studying small system collisions in high-energy experiments. A proton-proton collision can not reach local thermal equilibrium, however, an effective hydrodynamic description for the development of anisotropic flows may remain appropriate. These kind of questions have triggered in recent times a vast body of work \cite{Romatschke:2017ejr,Blaizot:2019scw,Berges:2020fwq}. 
%They have lead, in particular, to a paradigm shift, where the concept of hydrodynamization and the emergence of a hydrodynamic attractor \cite{PhysRevLett.115.072501}, first discussed in the context of holographic theories at strong coupling, replaces the traditional notion of local thermalization. 
Such approaches suggest, notably, that systems that do not reach local thermal equilibrium remain governed by effective constitutive relations formally equivalent to those of hydrodynamics. The possibility of studying these issues via the analysis of the elliptic flow of out-of-equilibrium mesoscopic gases would, thus, establish an even tighter connection between small systems collisions at the Large Hadron Collider and expanding ultracold clouds of few atoms in tabletop experiments.

\section{Acknowledgments}

We thank the members of the Collaborative Research Center \textit{``Isolated quantum systems and universality in extreme conditions''} (CRC 1225, ISOQUANT) for stimulating discussions related to this proposal. We are grateful, in particular, to Philipp Lunt, Kerthaan Subramanian, Carl Heintze, and Selim Jochim for detailed explanations related to the setup of cold atom experiments, as well as for a critical assessment of our idea which has largely helped develop this study. 
This work is supported by the Deutsche Forschungsgemeinschaft (DFG, German Research Foundation) under Germany's Excellence Strategy EXC 2181/1 - 390900948 (the Heidelberg STRUCTURES Excellence Cluster) and under 273811115 – SFB 1225 ISOQUANT as well as FL 736/3-1.
%This research is funded by DFG (German Research Foundation) – Project-ID 273811115 – SFB 1225 ISOQUANT.

\appendix

\section{Vanishing $v_2$ at high temperatures\label{app:v2AtHighTemp}}

We can argue that $v_2$ will vanish in the high temperature limit by introducing a cutoff for the quantum numbers,
\begin{equation}
    \lim_{T\rightarrow\infty} v_2 = \lim_{\beta\rightarrow 0} \lim_{\Lambda\rightarrow\infty} \frac{1}{Z_\Lambda}\sum_{n=0}^\Lambda \sum_{m=0}^\Lambda \expval{cos(2\phi_p)}_{n,m} e^{-\beta (n \omega_x + m \omega_y)} \, ,
\end{equation}
where $Z_\Lambda$ is the defined as in Eq. \eqref{eq:oneParticleDensityOp} with the sums only going to $\Lambda$.
Since the sums converge absolutely (the operator $|\cos(2\phi_p)|$ has a finite expectation value), we can exchange the limits,
\begin{equation}
\begin{split}
    \lim_{T\rightarrow\infty} v_2 &= \lim_{\Lambda\rightarrow\infty} \lim_{\beta\rightarrow 0} \frac{1}{Z_\Lambda}\sum_{n=0}^\Lambda \sum_{m=0}^\Lambda \expval{\cos(2\phi_p)}_{n,m} e^{-\beta (n \omega_x + m \omega_y)} \\
    &= \lim_{\Lambda\rightarrow\infty} \frac{1}{(\Lambda + 1)^2} \sum_{n=0}^\Lambda \sum_{m=0}^\Lambda \expval{\cos(2\phi_p)}_{n,m} \, .
\end{split}
\end{equation}
This is however just an average of $v_2$ over all states with $n\leq \Lambda$ and $m\leq \Lambda$.
In the limit $\Lambda\rightarrow\infty$ this is the same as averaging over all states which in turn is the same as the limit of infinite particle number $n\rightarrow\infty$ at zero temperature.
From the previous discussion we thus argue that
\begin{equation}
    \lim_{T\rightarrow\infty} v_2|_{N=1} = \lim_{N\rightarrow\infty} v_2|_{T=0} = 0 \, .
\end{equation}

Let $f$ be a bijection from $\mathbb{N}$ to $\mathbb{N}\times\mathbb{N}$.
Then the $T\rightarrow\infty$ limit of the elliptic flow can be written as
\begin{equation}
\begin{split}
    &\lim_{T\rightarrow\infty} v_2 |_{N=2} = 
    %\lim_{\beta\rightarrow 0} \frac{1}{2 Z_2} \sum_{n=0}^\infty \sum_{m=n+1}^\infty \left(\expval{\cos(2\phi_p)}_{f(n)} + \expval{\cos(2\phi_p)}_{f(m)} \right) e^{-\beta((f_x(n)+f_x(m))\omega_x + (f_y(n)+f_y(m))\omega_y)} \\
    %&= 
    \lim_{\beta\rightarrow 0} \lim_{\Lambda\rightarrow\infty} \frac{1}{2 Z_2^{(\Lambda)}} \sum_{n=0}^\Lambda \sum_{m=n+1}^\Lambda \left(\expval{\cos(2\phi_p)}_{f(n)} + \expval{\cos(2\phi_p)}_{f(m)} \right) e^{-\beta \sum_{i\in\{ x,y\}}(f_i(n)+f_i(m))\omega_i} \\
    &= \lim_{\Lambda\rightarrow\infty} \frac{1}{\Lambda (\Lambda + 1)} \sum_{n=0}^\Lambda \sum_{m=n+1}^\Lambda \left(\expval{\cos(2\phi_p)}_{f(n)} + \expval{\cos(2\phi_p)}_{f(m)} \right) \\
    &= \lim_{\Lambda\rightarrow\infty} \frac{1}{\Lambda (\Lambda + 1)} \left[ \sum_{n=0}^\Lambda (\Lambda - n) \expval{\cos(2\phi_p)}_{f(n)} + \sum_{n=0}^\Lambda \sum_{m=n+1}^\Lambda \expval{\cos(2\phi_p)}_{f(m)} \right] \\
    &= \lim_{\Lambda\rightarrow\infty} \frac{1}{\Lambda (\Lambda + 1)} \left[ \sum_{n=0}^\Lambda (\Lambda - n) \expval{\cos(2\phi_p)}_{f(n)} + \sum_{n=0}^\Lambda \sum_{m=0}^\Lambda \expval{\cos(2\phi_p)}_{f(m)} - \sum_{n=0}^\Lambda \sum_{m=0}^n \expval{\cos(2\phi_p)}_{f(m)} \right] \\
    &= \lim_{\Lambda\rightarrow\infty} \frac{1}{\Lambda (\Lambda + 1)} \left[ \sum_{n=0}^\Lambda (\Lambda - n) \expval{\cos(2\phi_p)}_{f(n)} + \Lambda \sum_{n=0}^\Lambda \expval{\cos(2\phi_p)}_{f(n)} - \sum_{n=0}^\Lambda (\Lambda - n + 1) \expval{\cos(2\phi_p)}_{f(n)} \right] \\
    &= \lim_{\Lambda\rightarrow\infty} \left( 1 - \frac{1}{\Lambda} \right) \frac{1}{(\Lambda + 1)} \sum_{n=0}^\Lambda \expval{\cos(2\phi_p)}_{f(n)} = \lim_{\Lambda\rightarrow\infty} \frac{1}{(\Lambda + 1)} \sum_{n=0}^\Lambda \expval{\cos(2\phi_p)}_{f(n)} \, .
\end{split}
\end{equation}
This is again just the average of $v_2$ over all states.
With this we get
\begin{equation}
    \lim_{T\rightarrow\infty} v_2|_{N=2} = \lim_{N\rightarrow\infty} v_2|_{T=0} = 0 \, .
    \label{eq:v2LargeTIsLargeN}
\end{equation}
Similar arguments can be made for higher particle numbers albeit with more complicated combinatorical considerations.
In general the argument is that (after exchanging the limits) the sums without considering the anti-symmetrization will give a contribution that is the desired average over all states with a prefactor of order one.
The part that is substracted to account for the overcounting will also be the state averaged value, but with prefactors that have negative powers of $\Lambda$.
In the limit $\Lambda\rightarrow\infty$, the first term dominates and relation \eqref{eq:v2LargeTIsLargeN} holds.
This argument holds for a bosonic gas as well where the sums in the above calculations start at $m=n$ instead of $m=n+1$ which only changes the structure of the subleading terms.

\printbibliography[heading=bibintoc]

\end{document}